\documentclass[useAMS,usenatbib,letters]{mn2e}
\usepackage{graphicx,color}
\usepackage[sf]{subfigure}
\usepackage{amssymb}
\usepackage{times}

\title[The nucleus of NGC 253 at parsec scales]{The nucleus of NGC 253 and its massive stellar clusters at parsec scales\thanks{Based on European Southern Observatory--Very Large Telescope programs \mbox{076.B-0493}, \mbox{074.A-9016} and \mbox{076.B-0656} and \emph{Hubble Space Telescope} programs U43A0103B and U31Q0101B}.}
\author[J.~A. Fern\'andez-Ontiveros, M.~A. Prieto and J.~A. Acosta-Pulido]{J.~A. Fern\'andez-Ontiveros$^{1}$\thanks{E-mail: \textsf{jafo@iac.es}}, M.~A. Prieto$^{1}$ and J.~A. Acosta-Pulido$^{1}$ \\
$^{1}$Instituto de Astrof\'isica de Canarias, V\'ia L\'actea s/n, La Laguna, E-38200, Spain}

\begin{document}

\date{Accepted 2008 October 10. Received 2008 October 10; in original form 2008 July 7}

\pagerange{\pageref{firstpage}--\pageref{lastpage}} \pubyear{2008}

\maketitle

\label{firstpage}

\begin{abstract}
Very Large Telescope adaptive optics images of NGC~253 with resolutions down to $200\, \rm{mas}$ resolve the central $300\, \rm{pc}$ of this galaxy in $\sim$37 infrared (IR) bright knots, a factor of 3 larger than previously reported, and extended diffuse emission. The angular resolution, comparable to that of available Very Large Array $2\, \rm{cm}$ maps, permits us a very accurate IR-radio registration. Eight radio sources are found to have an IR counterpart. The knots have H$\alpha$ equivalent width of about \mbox{80\,\AA}, sizes of $\sim$\,$3\, \rm{pc}$, magnitudes in $L$-band of about $12\, \rm{mag}$ and relatively high extinction, $A_V$\,$\sim$\,$7\, \rm{mag}$. Their spectral energy distributions (SEDs) look very similar, characterized by a maximum at $20\, \rm{\micron}$ and a gentle bump in the $1$--$2\, \rm{\micron}$ range. These features can be well reproduced by considering an important contribution of very young stellar objects to the IR, efficiently heating their dust envelope. The evidence indicates that these are young massive clusters bursting from their dust cocoons. A median SED of the knots is provided, which may represent one of the most genuine templates of an extragalactic circumnuclear star-forming region. The lack of any optical or IR counterpart for the previously identified radio core calls into question its supposed active nucleus nature. This source may instead represent a scaled up version of Sgr~A$^*$ at the Galactic Centre.
\end{abstract}
 


\section{Introduction}\label{intro}

NGC~253 is one of the nearest starburst galaxies, located at \mbox{$3.94 \pm 0.5\, \rm{Mpc}$} \citep{2003A&A...404...93K} in the Sculptor group. It is a nearly edge-on SAB(s)c galaxy hosting a number of young stellar clusters detected in the optical range in its nucleus, surrounded by large patches of dust \citep{2000MNRAS.312..689F}. A strong bipolar outflow cone is observed in ionized gas and X-ray emission \citep{2002ApJ...568..689S,2002ApJ...576L..19W}. At radio wavelengths (\mbox{$1.3$--$20\, \rm{cm}$}) \citet{1997ApJ...488..621U} identified more than $60$ individual sources along the north-east--south-west direction, probably associated with a $300\, \rm{pc}$ nuclear ring. Among them, a strong non-thermal compact source with brightness temperature of $T_{2\, \rm{cm}}$\,$>$\,$40\,000\, \rm{K}$ and spectral index of $\alpha_{1.3}^{3.6}$\,$=$\,$-0.3$ ($S_\nu$\,$\propto$\,$\nu^\alpha$) lies in its centre. In the infrared (IR), previous observations revealed the presence of bright hotspots in the nuclear starburst region \citep{1994ApJ...430L..33S}, with a bright source dominating the emission in this range \citep{2005A&A...438..803G}. The large change in morphology with wavelength led to different registrations in the literature \citep[see][and references therein]{2005A&A...438..803G}. As a result, different IR counterparts have been associated to the compact radio core. Adaptive optics IR data with full width at half-maximum (\textsf{FWHM})\,$=$\,$200\, \rm{mas}$ ($4\, \rm{pc}$) resolution, comparable to those in existing radio maps, allow us to make a very detailed registration of the central region. We resolve the inner $300\, \rm{pc}$ in 37 IR sources, finding radio counterparts for eight of them. This letter analyses their nature on the basis of high-spatial resolution spectral energy distributions (SEDs) and settles the location of the radio core, now absent at optical or IR wavelengths.


\section{Observations and the newly revealed morphology}\label{data}

Adaptive Optics images with the Very Large Telescope/\textsf{NaCo} (\textsf{VLT/NaCo}) [$0.0271\, \rm{arcsec\, pixel^{-1}}$, $28\times$\,$28\, \rm{arcsec^2}$ field of view (FOV)] were obtained during the nights of 2005 December 2 and 4 using the IR wavefront sensor (dichroic N90C10) and the brightest nuclear IR source for atmospheric correction. The filter set used was $J$ (total integration time: 600\,s; Strehl ratio: 3 per cent), $K_s$ (500\,s; 20 per cent), $L$ (4.375\,s; 40 per cent) and NB\_4.05 (8.75\,s, centred on Br$\alpha$; 40 per cent). Images with \textsf{VLT/VISIR} ($0.127\, \rm{arcsec\, pixel^{-1}}$, $32.3\times$\,$32.3\, \rm{arcsec^2}$ FOV) were taken using the $N$ (PAH2\_2, $\lambda\ 11.88\, \rm{\micron}$, $\Delta \lambda\ 0.37\, \rm{\micron}$; 2826\,s) and $Q$-bands (Q2, $\lambda\ 18.72\, \rm{\micron}$, $\Delta \lambda\ 0.88\, \rm{\micron}$; 6237\,s) the nights of 2004 December 1 and 2005 October 9. The achieved \textsf{FWHM}\footnote{Values measured on the most compact sources in each image.} resolutions were: $0.29\, \rm{arcsec}$ ($J$), $0.24\, \rm{arcsec}$ ($K_s$), $0.13\, \rm{arcsec}$ ($L$), $0.14\, \rm{arcsec}$ (NB\_4.05), $0.4\, \rm{arcsec}$ ($N$) and $0.74\, \rm{arcsec}$ ($Q$). Data reduction was done with the \textsf{ESO} packages \textsc{eclipse} and \textsc{esorex}, and includes sky subtraction, registration and combination of frames corresponding to each filter data set. Photometric calibration was based on standard stars observed next in time to the science targets. The IR set was complemented with \emph{Hubble Space Telescope} (\emph{HST}) images with the filters \emph{F555W} ($V$-band), \emph{F656N} (including H$\alpha$), \emph{F675W} ($R$-band), \emph{F814W} ($I$-band) and \emph{F850LP}; VLA maps from the literature at $2\, \rm{cm}$ (A configuration) by \citet{1985ApJ...299L..77T}, and at $1.3$, $2$, $3.6$, $6$ and $20\, \rm{cm}$ (A and B configurations) by \citet{1997ApJ...488..621U}; very long baseline interferometry (VLBI) radio maps at $13$ and $21\, \rm{cm}$ by \citet{2006AJ....132.1333L}.
\begin{figure*}
  \resizebox{\hsize}{!}{\includegraphics{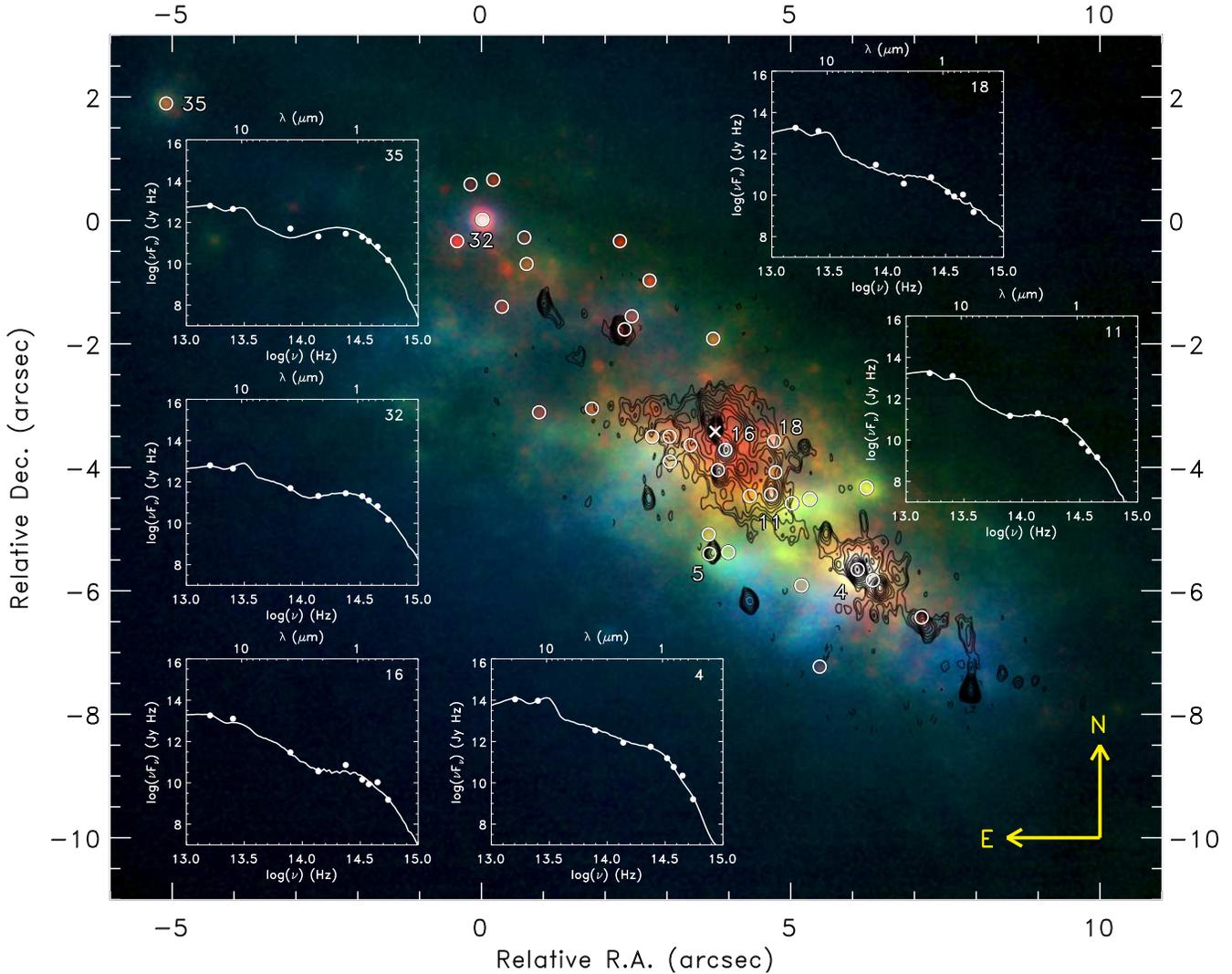}}
  \caption{\textsf{VLT} adaptive optics $J$ (blue), $K_s$ (green) and $L$ (red) bands colour composite image together with VLA $2\, \rm{cm}$ radio map contours \citep{1997ApJ...488..621U}. A cross marks the TH2 location, the assumed radio core, whereas circles mark the $L$-band sources found by \textsc{daophot} above 3$\sigma$ level. Insets show the SED and the best-fitting YSO model for a selection of regions (see Section~\ref{YSO}).}\label{color_radio}
\end{figure*}

Morphology of the central $\sim$\,$300\, \rm{pc}$ region from $1$ to $20\, \rm{\micron}$ resolves into multiple compact knots superposed on a diffuse emission component (Figs~\ref{color_radio} and \ref{Q}). They are mainly distributed along the disc of the galaxy in a ring-like structure, in line with \citet{2000MNRAS.312..689F}. However, whereas \citet{2000MNRAS.312..689F} resolve $\sim$\,$12$ sources, our high-spatial resolution data allow us to identify a factor of 3 more over the same region. The reduced extinction and higher spatial resolution achieved longward of $2.4\, \rm{\micron}$ led to the discovery of even further new sources without optical or near-IR (NIR) counterpart. The identification of the sources was done in the $L$-band image, as it has the highest contrast. Using the \textsc{daophot} finding algorithm \citep{1987PASP...99..191S}, a total of 37 knots with \textsf{FWHM}\,$\gtrsim$\,$0.13\, \rm{arcsec}$ and above 3$\sigma$ of the local background were identified within the central $\sim$\,$300\, \rm{pc}$ region, the majority being concentrated in the central $150\, \rm{pc}$. Most of the $L$-band knots are spatially resolved with a median size of \textsf{FWHM}\,$\sim$\,$3.3\, \rm{pc}$ ($0.17\, \rm{arcsec}$).

\subsection{Image alignment}\label{align}
\begin{figure}
  \resizebox{\hsize}{!}{\includegraphics{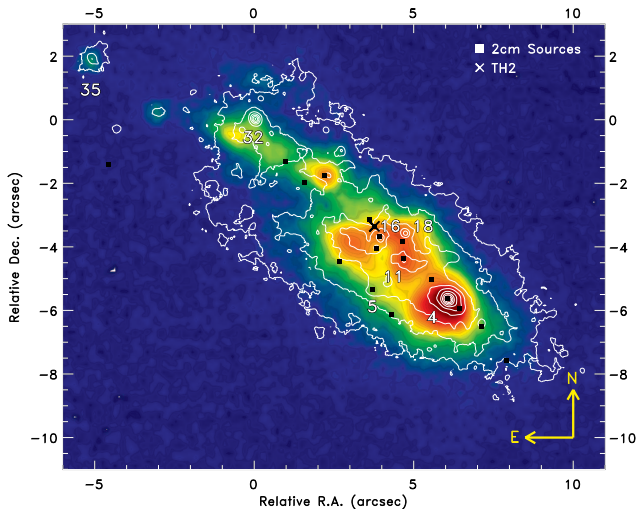}}
  \caption{$Q$-band image with $L$-band contours and VLA $2\, \rm{cm}$ source positions overlapped. The cross marks the location of TH2.}\label{Q}
\end{figure}

The IR images were registered using the brightest source, which is an outstanding feature at all wavelengths: knot \#4 in Fig.~\ref{color_radio}, labelled as M1 in \citet{2005A&A...438..803G}. In this way, $J$ and $K_s$-band knots are all found to have an $L$-band counterpart, although the converse is not true for a number of, probably very extinct, sources. Of the 37 sources identified in the $L$-band, 31 were detected in the \emph{HST/}Wide Field Planetary Camera 2 (\emph{HST}/WFPC2) $I$-band and 23 in the $V$-band. Most of the brightest or isolated knots in the $N$ and $Q$-bands were also found to have a counterpart in the $L$-band (Fig.~\ref{Q}). Further alignment between optical and IR images was based on the position of knot \#32, unambiguously identified at both spectral ranges. More than 10 NIR knots with a \emph{HST}/WFPC2 counterpart are found; furthermore, a good correspondence between the diffuse emission morphologies seen in both spectral ranges is found. The key alignment, however, is between radio and IR images. {\it The availability of the \textsf{VLT} adaptive optic $L$-band image with spatial resolution ($0.13\, \rm{arcsec}$) comparable to that of the $2\, \rm{cm}$ VLA map \citep[beam\,=\,$0.20 \times$\,$0.10\, \rm{arcsec^2}$;][]{1997ApJ...488..621U} allowed us to settle a very precise source identification in the NGC~253 nuclear region.} A first visual registration using point-like sources led to the identification of eight common knots. The final registration was obtained after distance minimization between $L$-band and $2\, \rm{cm}$ positions (Fig.~\ref{color_radio}). The achieved positional accuracy is $0.03\, \rm{arcsec}$.

The alignment was cross-checked with the $Q$-band (Fig.~\ref{Q}), in which it was found that the low brightness contours reproduce equivalent ones in the radio maps (e.g. when compared with $2\, \rm{cm}$ map in Fig.~\ref{color_radio}). Previous alignments are based on either an absolute or `blind' registration between radio and optical/IR \citep[e.g.][]{2000MNRAS.312..689F} or lower resolution MIR images \citep[e.g.][]{2005A&A...438..803G}. The strongest IR peak (knot \#4), which lacks a radio counterpart in those alignments, is now the counterpart of the thermal radio emitter TH7 \citep{1985ApJ...299L..77T}, also called \mbox{5.63--41.3} \citep{1997ApJ...488..621U}. Remarkably, the radio core TH2 or \mbox{5.79--39.1} is undetected at optical or IR wavelengths.


\section{The circumnuclear knots}\label{SED_models}

Photometry of each knot was extracted from a circular aperture (\textsf{r\,$=$\,FWHM} in each filter) centred on a common position for all filters, subtracting the mode from an annular concentric region (\textsf{r\,$\sim$\,[$2$--$4$]$\times$FWHM}) to account for the diffuse background emission. Magnitudes fall in the range $9.6 < L$-band\,$< 16.7\, \rm{mag}$.

Extinction values at each knot were derived from H$\alpha$/Br$\alpha$ estimate ratio \citep{1989agna.book.....O}. Line fluxes were determined, after continuum subtraction, from the narrow-band \emph{HST/F656N} and \textsf{VLT}/NB\_4.05 images, respectively. The continuum was inferred by interpolating between \emph{F555W} and \emph{F814W} images in the H$\alpha$ case; by scaling the $L$-band continuum to NB\_4.05 in the Br$\alpha$ case. In the latter, the scaling factor was determined by measuring the emission ratio between both filters in a number of interknot regions, and thus not affected by strong line emission. However, there is still a possible contamination by diffuse Br$\alpha$ emission which is the main uncertainty in the scaling factor. In such case, the Br$\alpha$ flux would be underestimated, and therefore the derived extinction would be a lower limit. We estimate the error magnitude to be $A_V$\,$\sim$\,$0.3\, \rm{mag}$, based on the scatter of the $L$-band/NB\_4.05 ratio values. Nevertheless, this error does not affect our results. The H$\alpha$ flux was corrected by the [N~{\small II}] contribution in the filter assuming that to be 40 per cent of the total \citep{1981PASP...93....5B}. Under these assumptions, a median $A_V$\,$\sim$\,$7.3\, \rm{mag}$ was found from all the knots. Reducing the [\textsc{N~ii}] contribution to 20 per cent lowers individual $A_V$ by $0.5\, \rm{mag}$. Their respective H$\alpha$ equivalent widths cluster about a median value of \mbox{\textsf{EW(H$\alpha$)}\,$=$\,$77$\,\AA}, with \mbox{$50$\,\AA} and \mbox{$150$\,\AA} as the first and third quartiles, respectively. Assuming instantaneous star formation \citep[e.g.][]{1999ApJS..123....3L}, those correspond to a median age of $\sim$\,$6.3\, \rm{Myr}$.

Table~\ref{phot} lists apparent magnitudes, extinctions ($A_V$), and \textsf{EW(H$\alpha$)} for a representative sample of knots with good signal-to-noise ratio (\textsf{S/N}). As $A_V$ values are derived assuming screen extinction, they must be considered as lower limits because of possible internal extinction. Fig.~\ref{color_radio} displays their SEDs from optical/\emph{HST} to IR/\textsf{VLT}. Their average size, $3.3\, \rm{pc}$, is comparable to that measured in M31 star clusters \citep{2004PASJ...56.1025K}. Further considering their IR magnitudes and extinctions, we conclude that they are probably young dust embedded clusters similar to those in NGC~5253 or the Antennae \citep{2004A&A...415..509V,2000ApJ...533L..57G}, which are highly obscured and exhibit a strong IR continuum.
\begin{table}
  \centering
  \caption{IR magnitudes, extinction (from H$\alpha /$Br$\alpha$ ratio) and \textsf{EW(H$\alpha$)} for a sample of regions in NGC~253.}\label{phot}
  \begin{tabular}{@{}cccccccc@{}}
    \hline
    Knot & $J$    & $K_s$   & $L$    & $N$   & $Q$   & $A_V$   & EW(H$\alpha$) \\
    (\#) & (mag)  & (mag)  & (mag)  & (mag) & (mag) & (mag)   & (\AA)         \\
    \hline
    4    & $14.6$ & $12.5$ & $9.6$  & $2.2$ & $0.4$ & $11.0$  & $61$          \\
    5    & $15.8$ & $14.4$ & $13.9$ & $5.6$ & $3.2$ & $5.4$   & $68$          \\
    11   & $16.7$ & $14.1$ & $13.0$ & $4.4$ & $2.4$ & $> 9.0$ & $< 200$       \\
    16   & $16.8$ & $16.0$ & $12.3$ & $4.4$ & $2.4$ & $10.3$  & $119$         \\
    18   & $16.3$ & $14.8$ & $12.7$ & $4.4$ & $2.0$ & $10.2$  & $24$          \\
    32   & $15.4$ & $14.1$ & $11.7$ & $5.5$ & $3.5$ & $8.7$   & $14$          \\
    35   & $16.8$ & $14.8$ & $14.0$ & $6.8$ & $4.6$ & $8.7$   & $49$          \\
    \hline
  \end{tabular}
\end{table}

\subsection{Average SED of the knots}\label{template}

 \begin{figure*}
  \centering
  \includegraphics[width=15cm]{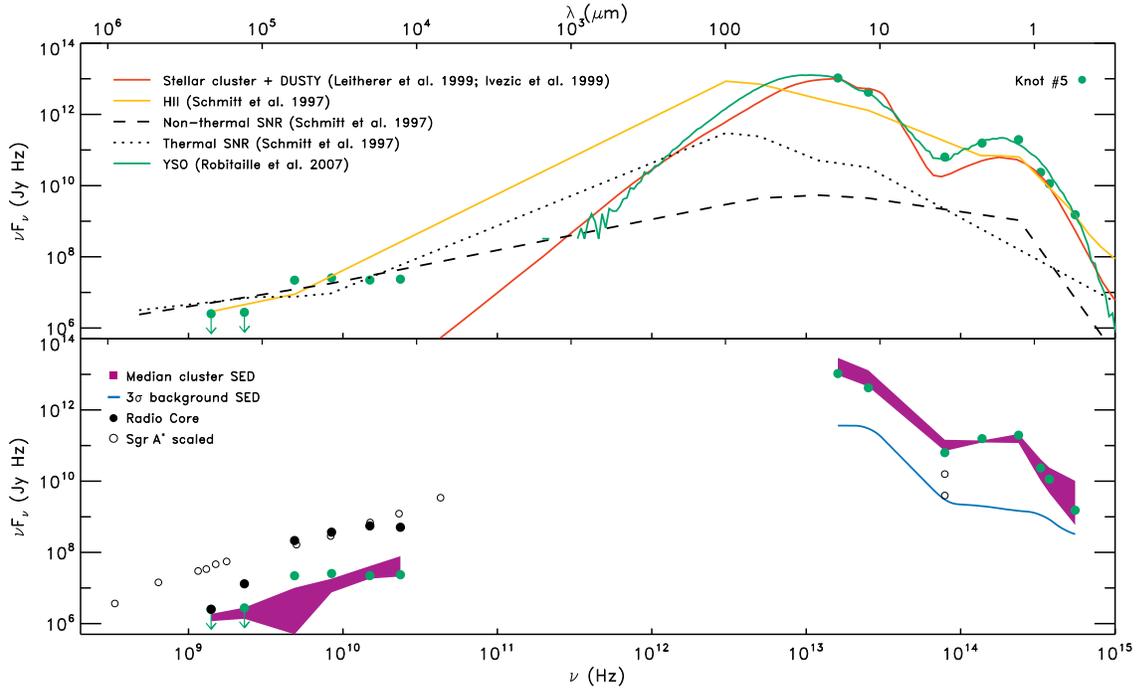}
  \caption{Top panel: SED templates for stellar cluster plus dust (red line), \textsc{H~ii} region (orange line), thermal SNR (dotted line), non-thermal SNR (dashed line) and YSO (green line) are compared with a representative SED (knot \#5, green circles). Bottom panel: knots median SED (shaded region) and the $3\sigma$ background (blue line) are compared with knot \#5 SED. Sgr~A$^*$ SED (open circles) scaled to radio core TH2 emission (solid circles).}\label{sed_completa}
\end{figure*}

SEDs were obtained for the 37 knots using our high-spatial resolution data. For the eight sources with radio counterparts, the SED extends from radio VLA-A to MIR subarcsec and NIR adaptive optics \textsf{VLT} to optical \emph{HST}. For the remaining sources, the SED is limited to the IR/\textsf{VLT} range, completed with optical/\emph{HST} data when detected. Most of the knots show very similar SEDs, characterized by a maximum in the MIR and a gentle `bump' at $\sim$\,$1\, \rm{\micron}$ (Fig.~\ref{color_radio}). To provide a genuine radio to optical SED template of a circumnuclear star-forming region, an average SED of all the knots was constructed. Each knot's SED was normalized to that of knot \#5 (green circles in Fig.~\ref{sed_completa}), using their respective median flux in the \mbox{$0.5$--$20\, \rm{\micron}$} range. The purple-coloured shaded region in Fig.~\ref{sed_completa} represents values between the first and third quartiles for each band scaled to knot \#5. The narrowness of the strip proves the similarity of individual SEDs. Background values measured around each knot were also averaged without normalizing. The SED for the $3\sigma$ averaged background is shown in Fig.~\ref{sed_completa} as a blue solid line. Comparing both -- median knot and background SEDs -- shows that the characteristic NIR bump in the \mbox{$1$--$2\, \rm{\micron}$} range is not a feature of the background emission but is intrinsic to the knots' SED, whereas the background SED is rather flat at these wavelengths (Fig.~\ref{sed_completa}). Similar NIR excesses have been detected in young stellar clusters like Henize~2--10 \citep{2005ApJ...631..252C} and NGC~1365 \citep{2008A&A...492....3G}, and are related to hot dust. Moreover, \citet{2008AJ....136.1415R} report an age--NIR excess correlation for the stellar clusters in SBS~0335--052. A similar trend is observed in NGC~253: redder colours ($I - J$\,$\sim$\,$4$) are found in the youngest clusters ($\lesssim$\,$4\, \rm{My}$) and bluer colours ($I - J$\,$\sim$\,$0$) in the oldest ($\sim$\,$10\, \rm{Myr}$). The ages are estimated from the \textsf{EW(H$\alpha$)} as indicated above (see Section~\ref{SED_models}). \emph{Thus, we interpret the NIR excess in these circumnuclear regions as hot dust emission related to young star-forming clusters.}

\subsection{Comparison with observational templates and models}
To investigate the nature of the knots, their SEDs were compared with theoretical models or observational templates of various types: an average \textsc{H~ii} region template, two types of supernova remnants, a stellar cluster model and a grid of templates for Young Stellar Objects (YSOs). For the sake of simplicity each of these templates was fitted to knot \#5 SED, using a $\chi^2$-minimization algorithm from which a scaling factor and a foreground extinction estimate $A_V$ are obtained \citep[galactic extinction law adopted from][]{1989ApJ...345..245C}. Fig.~\ref{sed_completa} compares the resulting fit to knot \#5 SED.

\subsubsection{\textsc{H~ii} region average template}\label{HII_region}
 \citet{1997AJ....114..592S} provide an average template from four well-known observed \textsc{H~ii} regions. The fitted template (orange line in Fig.~\ref{sed_completa}) shows a soft excess in the NIR ($\sim$\,$1\, \rm{\micron}$), and peaks at $\sim$\,$100\, \rm{\micron}$, providing a general account of the knot \#5 SED from radio to optical within one order of magnitude. In detail, its featureless shape in the $1$ to $20\, \rm{\micron}$ range largely departs from that of the knots. Given the particular knots' SED shape, the difference cannot be driven by extinction unless the applicable extinction law is very different from that in our neighbourhood.

\subsubsection{Supernova Remnant templates}\label{SNR}
The analysis of high-spatial resolution radio SED in NGC~253 suggests the presence of many supernova remnants (SNRs) in the nuclear region \citep{2006AJ....132.1333L}. Thus, the optical to radio SED of a non-thermal (Crab, dashed line) and a thermal (N~49, dotted line) SNR, both from \citet{1997AJ....114..592S}, were fitted to knot \#5 SED. Fig.~\ref{sed_completa} shows the large difference between these SEDs and that of the knots, in particular the inferred SNR IR fluxes are lower by several orders of magnitude with respect to those of the knots. The intrinsic shape of the original template is flatter than observed in knot \#5, which has a much higher IR/radio ratio. Additional extinction would only decrease this ratio, worsen the fit.\looseness-1

\subsubsection{Stellar cluster model plus dust}\label{SSC}
Knots' characteristics (Section~\ref{SED_models} above) point more in the direction of these being young star-forming regions. Accordingly, a stellar cluster template from \citet{1999ApJS..123....3L}, aged $6\, \rm{Myr}$ [in agreement with \textsf{EW(H$\alpha$)} age estimation] with instantaneous star formation and solar metallicity, was input to \textsc{dusty} \citep{1999astro.ph.10475I} to account for the dust emission. The best fit to knot \#5 SED assumes a spherical dust distribution with a temperature of $200\, \rm{K}$ on the shell inner boundary. $A_V$\,$=$\,$10\, \rm{mag}$ is needed to reproduce the SED optical range, a factor 2 larger than that estimated from H$\alpha$/Br$\alpha$ ratio. This may be related to the different assumptions on the extinction: \textsc{dusty} consider internal extinction, whereas we assume a simple foreground dust screen. However, the major discrepancy arises again in the NIR: the model underestimates the emission even if considering different metallicities, cluster ages, inner shell temperatures, grain sizes or dust density distributions. A possible contributing mechanism in this range is stochastic heating of small grains, an effect not included in \textsc{dusty}.

\subsubsection{Young Stellar Object models}\label{YSO}
Young stellar clusters appear to show SEDs very similar to those of YSO, suggesting that the early stages of their evolution may be parallel to those of massive stars \citep{2005IAUS..227..413J}. Motivated by this similarity, a comparison with YSO models was attempted. These are luminous IR sources formed by a protostar and an accretion disc, all enshrouded in a dense dust cloud \citep{1995PASJ...47..771N}. They show NIR excesses attributed to hot dust from the disc, a surplus that can be present in 30 per cent of the members of a young cluster \citep{2006AJ....131..951R}. Comparing with these models, a simplified scenario is adopted: YSOs are assumed as the main contributors to the knots' IR spectrum. Models were selected from the YSO grid provided by \citet{2007ApJS..169..328R}. Best fits to the knots' SED correspond to very young (\mbox{$0.1$--$3\, \rm{Myr}$}) luminous ($L_{Bol} > 10^2\, \rm{L_\odot}$) and massive stars ($M > 5\, \rm{M_\odot}$) seen at high inclination angles ($\sim$\,$80^\circ$). Extinction derived from the fits ranges from $A_V$\,$\sim$\,$5.0$ (knot \#5) to $11.6\, \rm{mag}$ (knot \#4), in fair agreement with those obtained from H$\alpha /$Br$\alpha$ ratios. If scaled the models to NGC~253 distance, a total of about $10^5$ YSOs per knot are required.

Overall, the optical to IR SED of individual knots are surprisingly well reproduced by these models (see Fig.~\ref{color_radio}), which naturally incorporates the hot dust contribution. The extinctions inferred from the fits being also in fair agreement with those measured from H$\alpha /$Br$\alpha$ ratios. The ages of the protostars are, however, smaller than those inferred from EW(H$\alpha$) (Table~\ref{phot}). {\it We believe that the NIR excess in NGC~253's circumnuclear regions represents the contribution of many YSOs bursting in their dust cocoons, and thus these are intense star-forming regions.}

\section{An active nucleus?}\label{nuclear_em}

The radio source TH2 \citep{1985ApJ...299L..77T} is the strongest ($21\, \rm{mJy}$ at $1.3\, \rm{cm}$), highest brightness-temperature ($T_{2\, \rm{cm}}$\,$>$\,$40\,000\, \rm{K}$) and most compact ($<$\,$2\, \rm{pc}$) source in the centre of this galaxy, and is thus assumed to be the radio core. The detection of a broad H$_2$O maser ($\Delta v$\,$\sim$\,$100\, \rm{km \, s^{-1}}$) and the large velocity gradient in the H92$\alpha$ radio-recombination line (\mbox{$\sim$\,$110\, \rm{km \, s^{-1} \, arcsec^{-1}}$}) suggest a dynamical mass of $M$\,$\approx$\,$7$\,$\times$\,$10^6\, \rm{M_\odot}$ \citep{1995PASJ...47..771N,2006ApJ...644..914R} at this position. Moreover, the H$92\alpha$ and H$75\alpha$ line emissions are only consistent with ionization produced by a stellar cluster younger than $10^5\, \rm{yr}$ or an AGN \citep{2002ApJ...574..701M}. The remarkable feature of this source is its lack of optical, NIR or MIR counterparts. The current IR-radio alignment reveals some diffuse IR emission at the position, yet, no point-like source commensurable with its radio strength is detected. The extinction at the location is comparable with the average in the region: $A_V$\,$\sim$\,$10\, \rm{mag}$ is measured at the closest knot, \#16 (Figs~\ref{color_radio} and \ref{Q}).

We compared TH2 with Sgr~A$^*$, a compact radio source with no optical counterpart, although much fainter than TH2. Fig.~\ref{sed_completa} (bottom panel) shows the radio spectrum of Sgr~A$^*$ \citep{2005ApJ...634L..49A} scaled to that of TH2 \citep{1995MNRAS.276.1373S,1997ApJ...488..621U}. Several IR flares have been detected in Sgr~A$^*$, with their flux limits in the $1$--$4\, \rm{mJy}$ range in $L$-band \citep{2004ApJ...601L.159G}. Both limits, without reddening correction and scaled by the same factor as the radio data, are also shown in Fig.~\ref{sed_completa}. They fall just $3\sigma$ above the NGC~253 diffuse background level, but are weaker than the $L$-band emission of knot \#5. {\it If the IR/radio ratio is expected to be higher in an AGN than in Sgr~A$^*$, we may then conclude that TH2 nature resembles closely that of Sgr~A$^*$ rather than an AGN.}\\


\section*{Acknowledgments}
\textsl{M. Orienti is acknowledged for reviewing the radio data. T. Robitaille for his comments on the YSO models. We also thank the referee for detailed comments which helped to improve the original manuscript. This work is partially funded by the Spanish MEC project AYA2006--09959.}


\bibliographystyle{mn2e}
\small
\bibliography{ngc253_NaCo.bib}

\label{lastpage}
\end{document}